\documentclass[pdflatex,sn-mathphys-num]{sn-jnl}
\usepackage{graphicx}%
\usepackage{epstopdf}
\usepackage{multirow}%
\usepackage{amsmath,amssymb,amsfonts}%
\usepackage{amsthm}%
\usepackage{mathrsfs}%
\usepackage[title]{appendix}%
\usepackage{xcolor}%
\usepackage{textcomp}%
\usepackage{manyfoot}%
\usepackage{booktabs}%
\usepackage{lmodern}
\usepackage{algorithm}%
\usepackage{algorithmicx}%
\usepackage{algpseudocode}%
\usepackage{listings}%
\usepackage{widetext}%
\theoremstyle{thmstyleone}%

\theoremstyle{thmstyletwo}%
\theoremstyle{thmstylethree}%
\begin{document}

\title[Article Title]{Near perfect noisy quantum teleportation by one-sided environment engineering}


\author[1]{\fnm{Md Manirul} \sur{Ali}}

\author[2]{\fnm{Sovik} \sur{Roy}}

\author[3, 4]{\fnm{Dipankar} \sur{Home}}

\affil[1]{\orgdiv{Centre for Quantum Science and Technology}, \orgname{Chennai Institute of Technology}, \orgaddress{\street{Sarathy Nagar, Kundrathur}, \city{Chennai}, \postcode{600069}, \country{India}}}

\affil[2]{\orgdiv{Department of Mathematics}, \orgname{Techno Main Salt Lake}, \orgaddress{\street{EM 4/1 Sector V}, Kolkata \postcode{700091}, \country{India}}}

\affil[3]{\orgname{Raman Research Institute (RRI)}, \orgaddress{\street{C. V. Raman Avenue}, \city{Bengaluru} \postcode{560080}, \country{India}}}

\affil[4]{\orgdiv{\orgname{Bose Institute}, \orgaddress{\street{EN Block, Sector V}, \city{Kolkata} \postcode{700091}, \country{India}}}}

\abstract{Achieving near-unity fidelity of Quantum Teleportation (QT) in a noisy environment is an essential requirement for its real-world
applications. Towards this goal, we devise a distinctive protocol based on the timing of the sender Alice’s Bell-basis measurement, synchronous
with the suitably engineered structure of noise in the receiver Bob’s wing, without requiring any knowledge or manipulation of Alice’s local noise.
For this purpose, to finally obtain the teleported states by performing the required unitary operations, Bob has to retain his qubits for \textit{only}
two particular Bell-basis outcomes communicated by Alice, whose corresponding reduced states have no dependence on the noise parameters in
Alice's wing. Such postselection is crucial for enhancing the fidelity of the teleported state towards its ideal limit, while ensuring its full independence
of Alice's noise. We formulate the specifics of our protocol in terms of a generic two-level quantum system, subjected to a general non-Markovian
spin-boson model for dephasing noise. The proposed scheme is illustrated by using any pure maximally/non-maximally entangled state as well
as a Werner-type mixed state as resource. In particular, using resource states having small values of entanglement measure, we show that the high
fidelity of QT is achievable in the presence of noise. Further, using our scheme, considering even the local regime of Werner states, wherein the
Bell-CHSH inequalities are not violated, we show that the QT fidelity in the presence of noise can be appreciably high.}


\maketitle

\section{Introduction}\label{sec1}

\noindent A remarkable quantum effect resourced by entanglement is Quantum Teleportation (QT) \cite{bennett1993teleporting,bouwmeester1997experimental,pirandola2015advances,hu2023progress,zeilinger2024dance},
one of the cornerstones of the {\it Second Quantum Revolution} \cite{dowling2003quantum,aspect2008introduction,jaeger2018second,deutsch2020harnessing,aspect2023second}.
This protocol leverages the non-classical nature of quantum entanglement for transferring an unknown quantum state from one system to another distant system, entailing a wide range of applications. In particular, QT has emerged as a key player in the current grand enterprise towards establishing a global quantum network, which hinges on long-distance quantum communication enabled by ground-to-satellite QT \cite{ren2017ground,li2022quantum}. Furthermore, quantum gate teleportation underpins the design of distributed quantum computing architecture as the basis for a scalable approach to fault-tolerant quantum computing \cite{main2025distributed}. Besides, in recent years, the study of QT using a variety of quantum states (e.g., involving multiple degrees of freedom, higher-dimensional quantum states) is gaining considerable importance in various contexts \cite{garcia2024quantum,wallnofer2020machine,xu2022machine,hu2023progress}. Here, it is also worth stressing the significant role of QT in developing quantum repeaters towards realizing an efficient quantum internet \cite{azuma2023quantum}.

\vskip 0.5cm

\noindent To recapitulate, the essence of QT lies in enabling the transfer of an unknown quantum state from, say, Alice's qubit ($A_1$) to a far-away qubit with Bob ($B$), using an entangled state preshared between the qubits ($A_2$ and $B$) possessed by Alice and Bob, respectively. For this purpose, the Bell-basis joint measurement (BM) has to be implemented by Alice on the qubits $A_1$ and $A_2$. This is followed by Alice sharing with Bob the outcomes of such measurements over a classical channel, and Bob's subsequent appropriate unitary rotations of the states. Now, in order to facilitate the efficient applications of such a protocol, it is crucial to ensure the high fidelity of QT in the presence of decohering noise in both the wings of Alice and Bob. Several studies have analyzed various aspects of QT under the Markovian and the non-Markovian decoherence dynamics \cite{horodecki1999general,badziag2000local,oh2002fidelity,bandyopadhyay2002origin,verstraete2003optimal,
ozdemir2007teleportation,bandyopadhyay2012optimal,taketani2012optimal,fortes2015fighting,fortes2016probabilistic,
laine2014nonlocal,im2021optimal,xu2024noise,zhang2024non,li2025one,ghosal2025repeater,hassan2026predicting}. In contrast to the Markovian open quantum systems exhibiting a memoryless evolution where the future state of a system depends only on its present state and not on its past history, the non-Markovian dynamics is crucially affected by the reservoir memory.  In this context, a critical open question concerning the implementation of QT in any noisy environment is to what extent the QT fidelity can be optimized by exploiting such non-Markovian memory effect in an operationally accessible way by fine-tuning a limited number of controllable parameters of the noisy environment.\\

\noindent A particularly interesting recent work tackling the above question has been the scheme due to Liu {\it et al.} \cite{liu2024overcoming}, essentially restricted to the photonic system. A variant of the standard QT protocol considered in their scheme is based on a hybrid entangled state shared between the photonic qubits possessed by Alice and Bob, respectively. Such a state is prepared by suitably coupling the photonic polarization states with the continuous frequency modes. An appropriately controlled local noise is introduced by Alice, while for achieving high QT fidelity, it is necessary for Bob to apply another local noise identical to that invoked by Alice. On the other hand, in the present paper, we formulate a scheme for mitigating noise in QT from an entirely different perspective.\\

\noindent
First, let us note two key aspects of the concept of fidelity as applied to QT, one is the fidelity
of the quantum channel \cite{horodecki1999general}, and the other one is the fidelity of the teleported state (\textit{FTS})
\cite{oh2002fidelity}. The fidelity of a quantum channel quantifies how well a quantum channel is capable of teleporting the state of a qubit, while the \textit{FTS} is quantified by the overlap between the state to be teleported and the actually teleported state. In this paper, whenever we talk about teleportation fidelity, we will mean fidelity of the teleported state (\textit{FTS}).
Next, we focus on outlining concisely the central features of our protocol whose resource can be any pure (maximally or non-maximally) entangled state or mixed entangled state, such as the Werner type state: \\

\noindent (a) In the QT scenario we consider (see Fig.~\ref{alice}), the time evolution is described under local dephasing noises acting in the respective wings of Alice and Bob up to the instant $t= \tau$ at which Alice's two qubits A1 and A2 are subjected to the Bell-basis joint measurement (BM). A salient aspect of our treatment is the adoption of a general framework for describing environment-induced decoherence based on the spin-boson model ($\grave{a}$ la Caldeira-Leggett) \cite{caldeira1983quantum}, applied for a generic two-level quantum system evolving in the presence of an archetypal non-Markovian dephasing noise. Further, an important feature of our formulated scheme is the occurrence of a decoherence-free-subspace (DFS) \cite{zanardi1997noiseless,lidar1998decoherence,lidar1999concatenating} in Alice’s wing. This pertains to the BM outcomes corresponding to
a particular pair of Bell states of Alice's two qubits ($A_1$ and $A_2$) which are not affected by the evolution under their common dephasing noise.
Consequently, the reduced density operators in Bob's wing corresponding to these Bell basis states arising from Alice's Bell measurement do not entail any
detrimental effect due to Alice's noisy environment. \\

\noindent On the other hand, the reduced density operator expressions corresponding to the other two BM outcomes communicated by Alice contain cumulative dephasing effects arising from the noises present in both the wings of Alice and Bob. Hence, towards maximizing the average FTS in our scheme, it is necessary to adopt the strategy of discarding Bob’s qubits for these two BM outcomes, and retain only those qubits of Bob corresponding to the former two BM outcomes. For this purpose, it suffices for Alice to inform Bob about the pair of BM outcomes to be discarded, without requiring to distinguish between the members of this pair. This would then require 1.5 bits of classical information to be conveyed from Alice to Bob in our reformulated QT scheme, unlike in the standard QT case where 2 classical bits are required to be sent.\\

\noindent (b) A key upshot of the above strategy is that the expression for the fidelity of the state teleported to Bob (FTS), averaged over the resource state parameters, turns out to be entirely independent of the noise parameters in Alice’s wing, and dependent only on the time-evolved value of the decoherence factor denoted by $b_\tau$ characterizing the local dephasing noise in Bob’s wing at the instant ($\tau$) Alice’s Bell-basis measurements are performed. Note that the decoherence factor $b_\tau$ is determined by the environmental noise parameters
in Bob's wing. Thus, here the crucial element lies in maximizing the average FTS by the optimum joint choice of the parameters $\tau$ and $b_\tau$. This operationally means that for ensuring high fidelity, Alice has to make an appropriate choice of the timing ($\tau$) of the Bell-basis measurement in her wing, depending upon her pre-shared knowledge of the spectral structure of Bob’s environmental noise, viz. the relevant decoherence parameter and its evolution in Bob’s wing. \\

\noindent Subsequently, in what follows, we proceed to elaborate upon the above characteristics in terms of the relevant mathematical details, starting from the treatment of the time evolutions in both the wings of Alice and Bob, up to the instant Alice’s Bell-basis measurement (BM) is performed. Then, by considering the reduced density operators in Bob’s wing corresponding to the different BM outcomes obtained by Alice, we outline the relevant specifics concerning the derivations of the expressions for average FTS for different resource states. This is done for both types of pure (maximally and non-maximally) entangled states, as well as using a mixed entangled Werner-type state. Subsequently, some representative quantitative estimates are given to highlight the efficacy of our scheme in achieving high values of FTS close to the maximum possible value by appropriately choosing the timings of BM in Alice’s wing, commensurate with the realisable values of the relevant noise parameters in Bob’s wing.

%%%%%%%%%%%%%%%%%%%%%%%%%%%%%%%%%%%%%%%%%%%%%%%%%%%%%%%%%%%%%%%%%%%%%%%%%%%%%%%%%%%%%%%%%%%%%%%%%%%%%%%%%%%%%%%%%%%%%%%%%%%%%%%%%%%%%%%%%%%%
\section{Results}\label{sec2}

\subsection{Specifics of our envisaged noisy teleportation scheme}\label{sec:quantumteleport}
\begin{figure}[h]
\centering
\includegraphics[width=10.7cm]{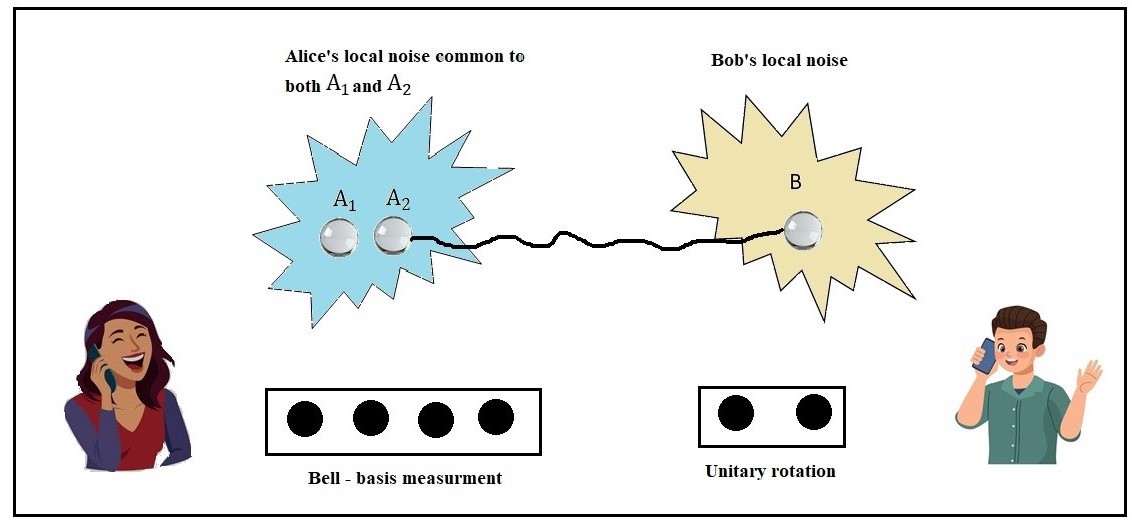}
\caption{The two qubits of Alice ($A_1$ and $A_2$) are interacting with a common dephasing environment. The qubit $B$ at Bob's
end evolves under another local dephasing noise, while $A_2$ and $B$ are initially entangled. Alice suitably times her Bell-basis
measurement (BM) on $A_1$ and $A_2$ in tune with Bob's noise parameters. Alice then communicates to Bob $1.5$ bits of information
regarding her BM outcomes. This enables Bob to appropriately select his qubits corresponding to a specific pair of BM outcomes, while
discarding his qubits for the other two outcomes. Subsequently, Bob performs suitable unitary operations to complete the teleportation
of state from $A_1$ to $B$.}
\label{alice}
\end{figure}

\noindent
The schematic of our protocol is shown in Fig. ~\ref{alice}. The state of Alice's qubit ($A_1$) is to be teleported
while the resource entangled state is shared between Alice's other qubit ($A_2$) and Bob's qubit ($B$).
The initial unknown state to be teleported  from $A_1$ to Bob is
\begin{eqnarray}
\label{instate}
\vert \psi_{in}\rangle = \alpha \vert \uparrow\rangle_{A_1} + \beta \vert \downarrow\rangle_{A_1}.
\end{eqnarray}
\noindent whose density operator representation is $\rho_{A_1}=\vert \psi_{in}\rangle\langle \psi_{in}\vert$ where
$\vert \alpha \vert ^2 + \vert \beta \vert ^2 = 1$. Clubbing the input state $\rho_{A_1}$ with the shared entangled
state $\rho_{A_2 B}$, the combined initial state at time $t=0$ can be written as $\rho_{A_1  A_2  B}$ ($=\rho_{A_1}
\otimes \rho_{A_2 B}$). The resource state $\rho_{A_2 B}$ shared by Alice and Bob may be a pure maximally or non-maximally
entangled state, or it can be a mixed entangled state, {\it e.g.} Werner-type state
\cite{werner1989quantum,peres1996separability,li2021information} which plays an important role in quantum
information theory as a paradigmatic family of mixed entangled states. \\

\noindent We now proceed to evaluate the average \textit{FTS} using pure and mixed entangled states respectively as resource for the noisy quantum teleportation protocol we have formulated.

\subsubsection{Evaluation of the average \textit{FTS} using maximally or non-maximally entangled pure state as a resource}

\noindent Here we consider the case of maximally or non-maximally entangled pure state as a resource given by
\begin{eqnarray}
\label{EntPure}
\vert \chi \rangle_{A_2 B} = \mu \vert \uparrow\rangle_{A_2}\vert \uparrow\rangle_{B} +
\lambda \vert \downarrow\rangle_{A_2}\vert \downarrow\rangle_{B}
\end{eqnarray}
where $\vert \mu \vert ^2 + \vert \lambda \vert ^2 = 1$. This resource state is shared between Alice ($A_2$) and Bob ($B$) at
the instant $t=0$. Further, Alice also has the unknown qubit state, which is to be teleported to Bob
({\it i.e.} $\vert \psi_{in}\rangle$), defined in Eq.~(\ref{instate}). The joint three-qubit state of Alice ($A_1,\:A_2$) and
Bob ($B$) at time $t=0$ is thus given by
\begin{eqnarray}
\label{jointstate1}
\vert \Psi\rangle_{A_1 A_2 B} &=& \vert \psi_{in} \rangle_{A_1} \otimes \vert \chi \rangle_{A_2 B} \\
\nonumber
&=& \alpha \mu~ \vert \uparrow\rangle_{A_1}\vert \uparrow\rangle_{A_2}\vert \uparrow\rangle_{B} +
\alpha \lambda~ \vert \uparrow\rangle_{A_1}\vert \downarrow\rangle_{A_2}\vert \downarrow\rangle_{B} \\
\nonumber
&& + \beta \mu~ \vert \downarrow\rangle_{A_1}\vert \uparrow\rangle_{A_2}\vert \uparrow\rangle_{B} +
\beta \lambda~ \vert \downarrow\rangle_{A_1}\vert \downarrow\rangle_{A2}\vert \downarrow\rangle_{B}.
\end{eqnarray}
\noindent Next, we analyze the dynamical evolution in noisy environment and derive the resulting expressions for the reduced density
operators in Bob's wing. We begin by noting the following aspects. Alice's local dephasing noisy environment has common effects
on her accessible two qubits $A_1$ and $A_2$. Bob's qubit $B$ evolves under a local dephasing noise from the time
$t=0$ to $t=\tau$. Our treatment is based on the spin-boson decoherence model applicable for a general class of two-level quantum systems in the presence of the non-Markovian dephasing noise. The dynamical evolution of this two-qubit system for the qubits $A_1$ and $A_2$ is governed by the Liouville-von Neumann master equation, described in the \textit{Methods} section. On the other hand,  the qubit $B$ of Bob evolves under a local dephasing noise, for which again we consider the spin-boson interaction Hamiltonian. Under this local dephasing noise, Bob's qubit $B$ evolves according to the relevant quantum master equation as discussed in the {\it Methods} section. \\

\noindent
After the noisy time evolutions from $t=0$ to $t=\tau$ in both the wings of Alice and Bob, Alice performs BM at time $t=\tau$, jointly on the qubits $A_1$ and $A_2$, and consequently the resulting three-qubit joint state of $A_1,\:A_2$ and $B$ becomes
\begin{eqnarray}
\rho_{A_1A_2B}(\tau) &=& \frac{1}{4}~ \vert \phi^{+} \rangle_{A_1A_2} \langle \phi^{+} \vert ~\otimes~
\rho^{\phi^{+}}_{B}(a_{\tau},b_{\tau}) + \frac{1}{4}~ \vert \phi^{-} \rangle_{A_1A_2} \langle \phi^{-} \vert
~\otimes~ \rho^{\phi^{-}}_{B}(a_{\tau},b_{\tau}) \nonumber \\
&+&  \frac{1}{4}~ \vert \psi^{+} \rangle_{A_1A_2} \langle \psi^{+} \vert ~\otimes~ \rho^{\psi^{+}}_{B}(b_{\tau})
+ \frac{1}{4}~ \vert \psi^{-} \rangle_{A_1A_2} \langle \psi^{-} \vert ~\otimes~ \rho^{\psi^{-}}_{B}(b_{\tau}),
\label{finalstate1}
\end{eqnarray}
where $a_\tau$ and $b_\tau$ are the time-evolved decoherence factors determined by the noise parameters in the respective
wings of Alice and Bob, while the orthonormal BM basis states are given by
\begin{eqnarray}
\label{bellbasis1}
\vert \phi^{\pm} \rangle_{A_1A_2} &=& \frac{1}{\sqrt{2}}\Big( \vert \uparrow \uparrow \rangle_{A_1A_2}
\pm \vert \downarrow \downarrow \rangle_{A_1A_2} \Big), \\
\vert \psi^{\pm} \rangle_{A_1A_2} &=& \frac{1}{\sqrt{2}}\Big( \vert \uparrow \downarrow \rangle_{A_1A_2}
\pm \vert \downarrow \uparrow \rangle_{A_1A_2} \Big).
\label{bellbasis2}
\end{eqnarray}
\noindent
Here we consider the specific values of the time-evolved decoherence factors at the instant $t=\tau$ when Alice
performs BM which results in the above joint state (\ref{finalstate1}) that represents a mixed ensemble characterized
by the following reduced density operators in Bob's wing, each occurring with probability $\frac{1}{4}$:
\begin{eqnarray}
\nonumber
\rho^{\phi^{+}}_{B}(a_{\tau},b_{\tau}) &=& 2 \vert \mu \vert^2 \vert \alpha \vert^{2} \vert \uparrow \rangle_{B}
\langle \uparrow \vert +
2 \mu \lambda^{\ast} \alpha \beta^{\ast} a_\tau b_{\tau}  \vert \uparrow \rangle_{B} \langle \downarrow \vert + \\
&&{} 2 \mu^{\ast} \lambda \alpha^{\ast} \beta a_{\tau}^{\ast} b_{\tau}^{\ast}  \vert \downarrow \rangle_{B}
\langle \uparrow \vert + 2 \vert \lambda \vert^2 \vert \beta\vert^2\vert \downarrow \rangle_{B} \langle \downarrow \vert,
\label{bm1}
\end{eqnarray}
\begin{eqnarray}
\nonumber
\rho^{\phi^{-}}_{B}(a_{\tau},b_{\tau}) &=& 2 \vert \mu \vert^2 \vert \alpha \vert^{2} \vert \uparrow \rangle_{B}
\langle \uparrow \vert -
2 \mu \lambda^{\ast} \alpha \beta^{\ast} a_{\tau} b_{\tau}  \vert \uparrow \rangle_{B} \langle \downarrow \vert - \\
&&{} 2 \mu^{\ast} \lambda \alpha^{\ast} \beta a_{\tau}^{\ast} b_{\tau}^{\ast} \vert \downarrow \rangle_{B} \langle \uparrow \vert +
2 \vert \lambda \vert^2 \vert \beta\vert^2\vert \downarrow \rangle_{B} \langle \downarrow \vert,
\label{bm2}
\end{eqnarray}
\begin{eqnarray}
\nonumber
\rho^{\psi^{+}}_{B}(b_{\tau}) &=& 2 \vert \lambda \vert^2 \vert \alpha\vert^{2}\vert \downarrow \rangle_{B} \langle \downarrow\vert +
2 \mu^{\ast} \lambda \alpha \beta^{\ast} b_{\tau}^{\ast} \vert \downarrow \rangle_{B} \langle \uparrow \vert + \\
&&{} 2 \mu \lambda^{\ast} \alpha^{\ast}\beta~b_{\tau} \vert \uparrow \rangle_{B} \langle \downarrow \vert +
2 \vert \mu \vert^2 \vert \beta \vert^2 \vert \uparrow \rangle_{B} \langle \uparrow \vert,
\label{bm3}
\end{eqnarray}
\begin{eqnarray}
\nonumber
\rho^{\psi^{-}}_{B}(b_{\tau}) &=& 2 \vert \lambda \vert^2 \vert \alpha\vert^2 \vert \downarrow \rangle_{B}\langle \downarrow\vert -
2 \mu^{\ast} \lambda \alpha \beta^{\ast} b_{\tau}^{\ast} \vert \downarrow \rangle_{B} \langle \uparrow \vert - \\
&&{} 2 \mu \lambda^{\ast} \alpha^{\ast}\beta~b_{\tau} \vert \uparrow \rangle_{B} \langle \downarrow \vert +
2 \vert \mu \vert^2 \vert \beta \vert^2 \vert \uparrow \rangle_{B} \langle \uparrow \vert.
\label{bm4}
\end{eqnarray}
\textbf{\it Implications of the above forms of the reduced density operators:} At this stage, it is important to note that for the two outcomes of Alice's BM, corresponding to the BM basis states $\vert \phi^{+} \rangle_{A_1A_2}$ and $\vert \phi^{-} \rangle_{A_1A_2}$, the relevant reduced states $\rho^{\phi^{+}}_{B}(a_{\tau},b_{\tau})$ and $\rho^{\phi^{-}}_{B}(a_{\tau},b_{\tau})$
in Bob's wing, have strong dephasing, as evidenced by Eqs.~(\ref{bm1}) and (\ref{bm2}). This is because the noises arising from
the environments of Alice ($a_{\tau}$) and Bob ($b_{\tau}$) have a cumulative effect. As easily seen from the expressions of the
density matrices (\ref{bm1}) and (\ref{bm2}), the off-diagonal elements of $\rho^{\phi^{+}}_{B}(a_{\tau},b_{\tau})$ and $\rho^{\phi^{-}}_{B}(a_{\tau},b_{\tau})$ loose coherence rapidly due to the effects of combined dephasing.
Hence, to counteract this effect, the strategy we devise is that Bob will \textit{discard} his qubit states for these two outcomes
and \textit{retain} his qubits only for the other two outcomes of Alice's BM results communicated to him. These two retained
outcomes pertain to $\vert \psi^{+} \rangle_{A_1A_2}$ and $\vert \psi^{-} \rangle_{A_1A_2}$ whose corresponding reduced density operators in Bob's wing $\rho^{\psi^{+}}_{B}(b_{\tau})$ and $\rho^{\psi^{-}}_{B}(b_{\tau})$ given by Eqs.~(\ref{bm3}) and
(\ref{bm4}) do not reflect any dephasing effect from Alice's side. This key feature of circumventing any dependence on noise in Alice's wing arises from the fact that the two states  $\vert \psi^{+} \rangle_{A_1A_2}$ and $\vert \psi^{-} \rangle_{A_1A_2}$ in Alice's wing are degenerate eigenstates of a collective spin operator of $A_1$ and $A_2$, satisfying the condition for the existence of a
\textit{decoherence-free subspace} \cite{zanardi1997noiseless,lidar1998decoherence,lidar1999concatenating}. \\

\noindent Next, in the final stage, Bob applies appropriate unitary rotations to the states $\rho^{\psi^{+}}_{B}(b_{\tau})$ and $\rho^{\psi^{-}}_{B}(b_{\tau})$ based on the information received from Alice about her BM outcomes. For instance, for the specific outcome corresponding to $\vert \psi^{+} \rangle_{A_1A_2}$ communicated by Alice, Bob can apply the unitary operator $U_{x}=\sigma_{x}$ to transform his reduced qubit state to the following form, which is given by
\begin{eqnarray}
\nonumber
\rho^{\psi^{+}}_{out}(b_{\tau}) &=& U_{x}~\rho^{\psi^{+}}_{B}(b_{\tau})~U_{x}^{\dagger} \\
\nonumber
&=& 2 \vert \lambda \vert^2 \vert \alpha \vert^{2} \vert \uparrow \rangle_{B} \langle \uparrow \vert +
2 \mu^{\ast} \lambda \alpha \beta^{\ast} b_{\tau}^{\ast} \vert \uparrow \rangle_{B} \langle \downarrow \vert + \\
&&{} 2 \mu \lambda^{\ast} \alpha^{\ast} \beta b_{\tau} \vert \downarrow \rangle_{B} \langle \uparrow \vert
+ 2 \vert \mu \vert^2 \vert \beta\vert^2 \vert \downarrow \rangle_{B} \langle \downarrow \vert.
\label{Ux1}
\end{eqnarray}
On the other hand, for the specific outcome corresponding to $\vert \psi^{-} \rangle_{A_1A_2}$
communicated by Alice, Bob can apply the unitary operator $U_{y}=i\sigma_{y}$ to transform
his reduced qubit state to the form given by
\begin{eqnarray}
\rho^{\psi^{-}}_{out}(b_{\tau}) = U_{y}~\rho^{\psi^{-}}_{B}(b_{\tau})~U_{y}^{\dagger} = \rho^{\psi^{+}}_{out}(b_{\tau}).
\label{Uy1}
\end{eqnarray}

\noindent
\textbf{\it Deriving average \textit{FTS}:} From the expressions for the reduced density operators of Bob given by the above Eqs.~(\ref{Ux1}) and (\ref{Uy1}),
we now proceed to estimate the average \textit{FTS} in our
protocol for which Bob postselects his qubits corresponding to these density operators for completing the teleportation process. We first calculate the teleportation fidelity as follows:
\begin{eqnarray}
\nonumber
&& f(\theta,\phi, b_{\tau}) = \langle \psi_{in} \vert \rho_{out}^{\psi^{\pm}} (b_{\tau}) \vert \psi_{in} \rangle
= 2 \vert \lambda \vert^2 \vert \alpha \vert^4 + 2 \vert \mu \vert^2 \vert \beta \vert^4
+ 2 \vert \alpha \vert^2 \vert \beta \vert^2 \Big(\mu \lambda^{\ast} b_{\tau}
+ \mu^{\ast} \lambda b_{\tau}^{\ast} \Big) \\
{}&& = 2 \vert \lambda \vert^2 \cos^4 (\theta/2) + 2 \vert \mu \vert^2 \sin^4 (\theta/2)
 + 2 \sin^2 (\theta/2) \cos^2 (\theta/2) \Big(\mu \lambda^{\ast} b_{\tau}
+ \mu^{\ast} \lambda b_{\tau}^{\ast} \Big)
\label{purefid2}
\end{eqnarray}
where the unknown qubit state to be teleported given by $\vert \psi_{in}\rangle$ is expressed in terms of the Bloch sphere parameters
$\alpha=\cos(\theta/2)$, and $\beta=\sin(\theta/2)e^{i\phi}$. Since $\vert \psi_{in}\rangle$ is unknown, it is relevant to estimate the average \textit{FTS} \cite{oh2002fidelity} by integrating over the ranges of $\theta$ and $\phi$ in the qubit Bloch sphere. Finally, we obtain the average fidelity as
\begin{eqnarray}
\label{avtelepfidpure}
{\cal F}_P(b_{\tau}) = \frac{1}{4\pi}\int_{0}^{\pi} \!\!\!\!d\theta
\int_{0}^{2\pi} \!\!\!\!d\phi ~f(\theta,\phi,b_{\tau}) \sin(\theta)
= \frac{2}{3} + \frac{1}{3} \Big( \mu \lambda^{\ast} b_{\tau} + \mu^{\ast} \lambda b_{\tau}^{\ast} \Big).
\end{eqnarray}

\noindent
Here $\mu$ and $\lambda$ are the parameters of the shared entangled state $\vert \chi \rangle_{A_2 B}$
(given by Eq.~(~\ref{EntPure})), and the value of the time-evolved decoherence factor $b_{\tau}$ at the instant
$\tau$ when BM is performed by Alice is determined by the spectral structure of the noisy environment in Bob's
wing $B$. From Eq.~(\ref{avtelepfidpure}) it is clear that there is no dependence of the average FTS on the noisy
effects from Alice's wing; in other words, given any entangled state as resource, the average FTS is essentially
determined by the nature of the noise in Bob's wing.  \\

\noindent \textit{Effect of noise in Bob's wing and optimisation of the average FTS:} Now, in order to enable
effective optimisation of the average FTS by appropriately choosing the timing ($\tau$) of Alice's BM, depending
upon the noise parameters in Bob's wing, it is necessary to assume a specific form of the spectral density for Bob's
noisy environment and derive detailed expression for the decoherence factor $b_{\tau}$. For this, we proceed by
considering an Ohmic spectral density \cite{leggett1987dynamics} for the noisy environment in Bob's wing, given by
\begin{eqnarray}
\label{Jomega}
{\cal J} (\omega) = \gamma \omega \exp\Big(-\frac{\omega}{\Lambda}\Big)
\end{eqnarray}
where $\gamma$ represents the system-environment coupling strength between the system and the noisy environment, and $\Lambda$ is the cutoff frequency that determines the range of allowed frequencies in the environmental reservoir model. The noise reservoir in Bob's wing is considered to be at an extremely low temperature ($k_{B}T\sim 0$) with $k_{B}$ being the Boltzmann constant and $T$ is the temperature of the bath, resulting in a time-dependent decay  \cite{breuer2002theory} of the off-diagonal elements of the single qubit reduced density operator in Bob's wing, given by the following time-dependent decay rate
\begin{eqnarray}
\label{deco01}
B(t) = 4 \gamma \frac{\Lambda^2 t}{1+\Lambda^2 t^2}.
\end{eqnarray}

\noindent
Now, using the above expression for the time-dependent decay rate, the time evolved decoherence factor in Bob's wing at the
instant $t=\tau$ is given by
\begin{eqnarray}
b_{\tau} = e^{-i \omega_{0}\tau - \int_{0}^{\tau} B(t) dt}
= e^{-i \omega_{0}\tau - 2 \gamma \ln (1+\Lambda^2 \tau^2)}.
\label{deco02}
\end{eqnarray}
It is thus seen from Eq. ~(\ref{deco02}) that the decoherence factor $b_\tau$ is essentially determined by the
environmental noise parameters $\gamma$, $\Lambda$ in Bob's wing, while $\omega_0$ is the qubit
transition frequency. Here, for completeness, note that the general form of the time-dependent decay rate
$B(t)$ occurring in the quantum master equation (\ref{masterB}) is given by Eq.~(\ref{decoB}) of
{\it Methods} Section. \\

\noindent
\textit{Quantitative study towards maximising the average FTS:} Let us consider as resource a non-maximally entangled
pure state $\vert \chi \rangle_{A_2B}$ shared between Alice ($A_2$) and Bob ($B$) at $t=0$. Within the framework of
our noisy teleportation protocol, we substitute the expression of the decoherence factor $b_{\tau}$ from Eq.~(\ref{deco02}) into
Eq.~(\ref{avtelepfidpure}) to obtain the average fidelity of the teleported state as
\begin{eqnarray}
\label{purefid7}
{\cal F}_P(b_{\tau}) = \frac{2}{3} + \frac{1}{3}~\mathcal{C}_{p} \cos(\omega_0 \tau) e^{-2 \gamma \ln(1+ \Lambda^2 \tau^2)}
\end{eqnarray}
where $\mathcal{C}_{p} =2 \mu \lambda$ is the entanglement measure of the pure state $\vert \chi \rangle_{A_2B}$ given by
Wootters' formula for concurrence \cite{wootters2001entanglement}, while the parameters of the resource pure state given by
$\mu$ and $\lambda$ are taken to be real. The results for the maximally entangled channel can be easily obtained if we choose
$\mu=\lambda=1/\sqrt{2}$ for which $\mathcal{C}_{p}=1$. \\

\begin{figure}[h]
\centering
\includegraphics[width=\columnwidth]{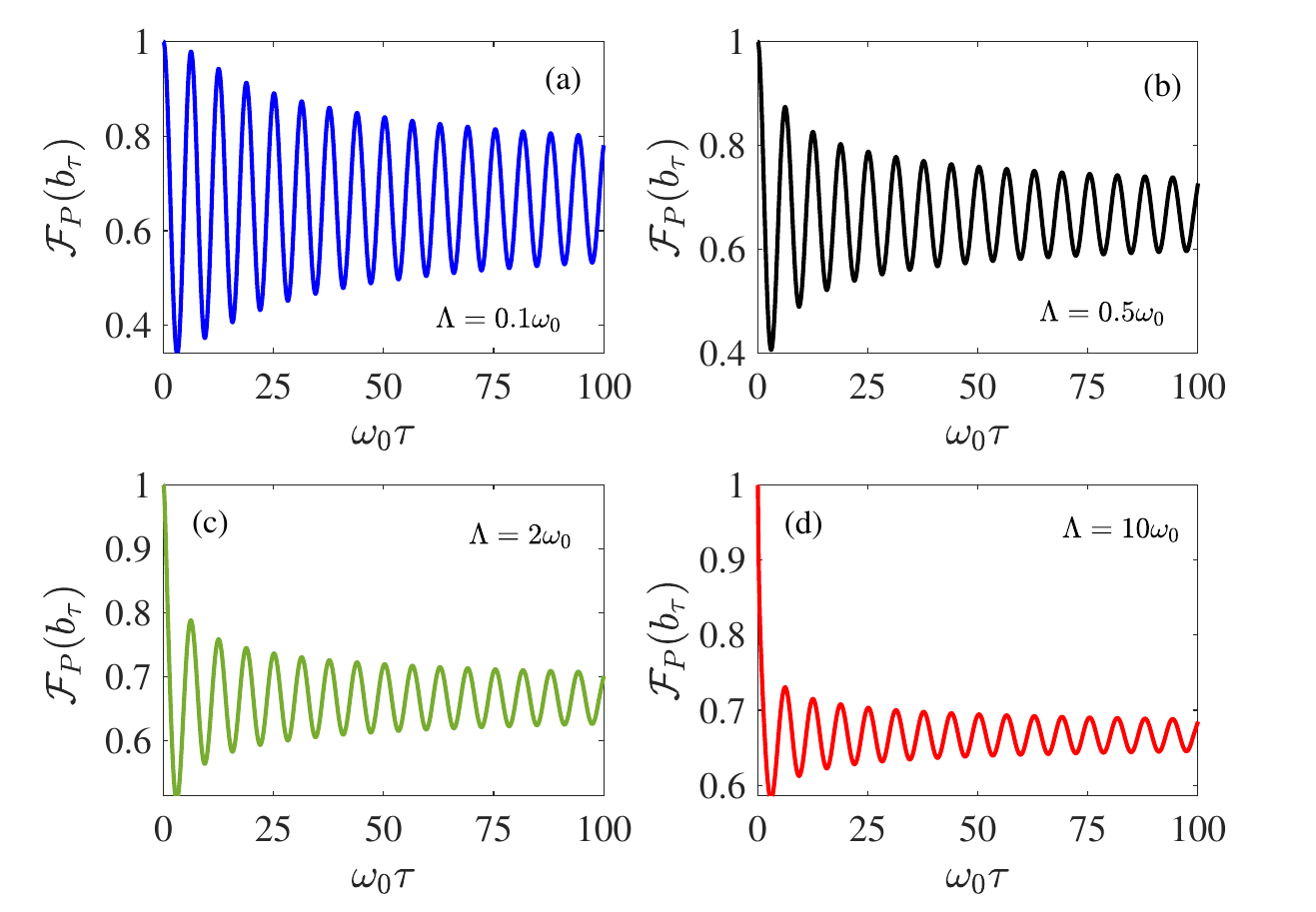}
\caption{\label{purefid01} We present the average fidelity of the teleported state ${\cal F}_P(b_{\tau})$
as a function of $\tau$-the Bell-basis measurement timing in Alice's wing. The concurrence of the shared
pure maximally entangled state is $\mathcal{C}_{p}=1$. The system-environment coupling strength in Bob's wing
is chosen as $\gamma=0.1\omega_0$. The panels (a)-(d) illustrate the results for various cutoff frequencies
pertaining to the different realizations of the dephasing noise: (a) $\Lambda=0.1\omega_0$, (b) $\Lambda
=0.5\omega_0$, (c) $\Lambda=2\omega_0$, and (d) $\Lambda=10\omega_0$, respectively.}
\end{figure}
\begin{figure}[h]
\centering
\includegraphics[width=\columnwidth]{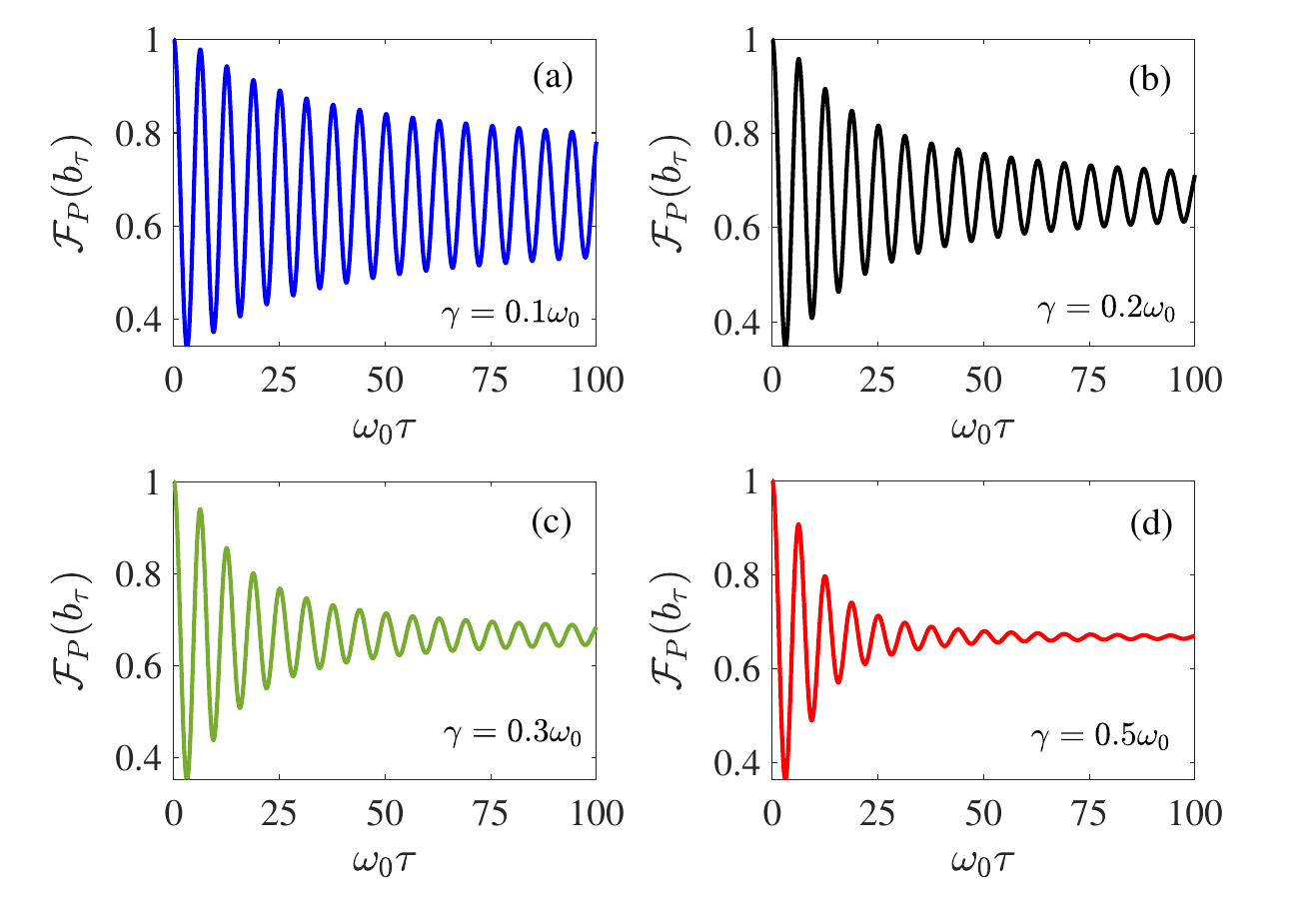}
\caption{\label{purefid02} We present the average fidelity of the teleported state ${\cal F}_P(b{\tau})$ as
a function of the Bell-basis measurement timing $\tau$ in Alice’s wing, for different system-environment
coupling strengths $\gamma$. The concurrence of the shared pure maximally entangled state is
$\mathcal{C}_{p}=1$, and the cutoff frequency of the noisy environment is chosen as $\Lambda=0.1\omega_0$.
The panels (a)-(d) display the results for increasing coupling strengths: (a) $\gamma=0.1\omega_0$, (b)
$\gamma=0.2\omega_0$, (c) $\gamma=0.3\omega_0$, and (d) $\gamma=0.5\omega_0$, respectively.}
\end{figure}

\noindent
In Fig.~\ref{purefid01}, we plot the average fidelity of the teleported state ${\cal F}_P(b_{\tau})$ as a function of Alice’s
measurement timing $\tau$ for different values of the cutoff frequency $\Lambda$ pertaining to the dephasing noise
model in Bob's wing. According to the formulation of our QT protocol, the values of the environmental noise model
parameters $\Lambda$ and $\gamma$ in Bob's wing are fixed in advance, and shared with Alice. As shown in
Fig.~\ref{purefid01}, the system-environment coupling strength is chosen to be $\gamma=0.1\omega_0$.
The cutoff frequencies of Bob’s environmental noise are chosen as (a) $\Lambda=0.1\omega_0$,
(b) $\Lambda=0.5\omega_0$, (c) $\Lambda=2\omega_0$, and (d) $\Lambda=10\omega_0$.
Based on this prior knowledge, Alice can select her measurement timing $t=\tau$ to optimize the average
fidelity ${\cal F}_P(b_{\tau})$ given by Eq.~(\ref{purefid7}). Thus, the choice of $\tau$ in Alice’s wing
depends critically on the pre-shared information about Bob’s noise parameters. Note that the parameter
values used above for illustration are adopted from relevant non-Markovian dephasing noise models
discussed in Refs. \cite{goan2010non,chen2014investigating,addis2014coherence}. \\

\noindent From Fig.~\ref{purefid01}(a), it is seen that the average fidelity ${\cal F}_P(b_{\tau})$ exhibits
oscillatory behavior with $\tau$ and decays slowly over time. For a fixed set of noise parameters
$\Lambda$, $\gamma$, the local maxima of ${\cal F}_P(b_{\tau})$ occur for various values of
$\omega_0 \tau = 2n\pi$, where $n=1,2,3,\ldots$. Therefore, to achieve a higher value of the average
teleportation fidelity, Alice needs to perform her Bell-basis measurement sufficient early. Specifically,
Fig.~\ref{purefid01}(a) shows that the maximum average fidelity can exceed $0.9$ if Alice performs
her measurement within the range $0 < \omega_0 \tau < 8\pi$. As the cutoff frequency $\Lambda$
increases, a faster decay of the average fidelity is observed, as illustrated in Fig.~\ref{purefid01}(b)-(d),
indicating stronger degradation of teleportation performance for wider environment bandwidths in Bob's wing. \\

\noindent In Fig.~\ref{purefid02}, we plot the average fidelity ${\cal F}_P(b_{\tau})$ as a function of Alice’s
measurement timing $\tau$ for different system-environment coupling strengths $\gamma$, while keeping the
cutoff frequency fixed at $\Lambda=0.1\omega_0$. The coupling strengths are chosen as (a) $\gamma=0.1\omega_0$, (b) $\gamma=0.2\omega_0$, (c) $\gamma=0.3\omega_0$, and (d) $\gamma=0.5\omega_0$. The fidelity exhibits
oscillatory behavior with maxima occurring near $\omega_0\tau=2n\pi$. As the coupling strength increases, the
oscillations damp more rapidly, leading to a faster decay of ${\cal F}_P(b_{\tau})$. Consequently, the time window
over which high teleportation fidelity is achievable becomes progressively narrower for larger $\gamma$. This
indicates that stronger system-environment coupling enhances decoherence, requiring Alice to perform the Bell-basis
measurement at earlier times to optimize the teleportation performance. \\

\noindent
Table~$1$ presents the average fidelity ${\cal F}_P(b_{\tau})$ of the teleported state for different values of the
concurrence $\mathcal{C}_{p}$ of the pure entangled state (\ref{EntPure}), the values being varied using the
parameters $\mu$ and $\lambda$, while the noise parameters in Bob's wing are fixed at $\gamma=0.1\omega_0$ and
$\Lambda=0.2\omega_0$. Alice performs the Bell-basis measurement at the optimal timing $\omega_0\tau=2\pi$,
determined by these pre-shared values of Bob's noise model parameters. We also examine the dependence of the average FTS on Bell nonlocality of the resource state given by $\vert \chi \rangle_{A_{2}B}$. For each value
of $\mathcal{C}_{p}$, the corresponding maximum violation of the Bell-CHSH inequality given by $\langle \mathcal{B}\rangle_{\max}^{|\chi\rangle}$ for
the state $|\chi\rangle_{A_2B}$ is computed (see Methods). For a given concurrence and its associated Bell violation,
we compute the optimal average teleportation fidelity achievable for a fixed set of noise parameters. The results show
a monotonic increase of ${\cal F}_P(b_{\tau})$ with $\mathcal{C}_{p}$ and $\langle \mathcal{B}\rangle_{\max}^{|\chi\rangle}$.
It is seen that by appropriately tuning the measurement timing $\tau$ in accordance with Bob’s noise parameters $(\gamma,\Lambda)$,
different fidelity values can be obtained. A particular point worth noting here is that even for relatively smaller values of concurrence and violations of the Bell-CHSH inequality, appreciable values of average teleportation fidelity within the range $0.7 - 0.8$ can be achieved in the presence of noise. \\

\begin{table}[h]
\caption{Average teleportation fidelity ${\cal F}_P(b_{\tau})$ for different values of concurrence $\mathcal{C}_p$ and maximum
Bell-CHSH violation $\mathcal \langle B \rangle_{max}^{\vert \chi\rangle}$}\label{tab1}
\begin{tabular}{@{}lllllllllll@{}}
\toprule
$\mathcal C_p$  & $0.1$ & $0.2$ & $0.3$ & $0.4$ & $0.5$ & $0.6$ & $0.7$ & $0.8$ & $0.9$ & $1.0$ \\
\midrule
$\mathcal \langle B\rangle_{max}^{\vert \chi\rangle}$  & $2.09$ & $2.19$ & $2.28$ & $2.36$ & $2.44$ & $2.52$ & $2.60$ & $2.68$ & $2.75$ & $2\sqrt{2}$\\
\midrule
${\cal F}_P(b_{\tau})$ & $0.69$ & $0.72$ & $0.74$ & $0.77$ & $0.80$ & $0.83$ & $0.85$ & $0.88$ & $0.91$ & $0.94$\\
\end{tabular}
\end{table}

\subsubsection{Noisy teleportation and average \textit{FTS} using mixed entangled Werner type state as resource}

\noindent
Next, we consider the resource state shared by Alice ($A_2$) and Bob ($B$) to be a mixed entangled Werner-type state \cite{bouwmeester2000physics}, given by
\begin{eqnarray}
\label{EntMix}
\rho_{A_2 B}^{w} = p~ \vert \phi^{+} \rangle_{A_2 B} \langle \phi^{+} \vert
+ \frac{1}{4} \left(1-p \right) \left( I \otimes I  \right)_{A_2 B},
\end{eqnarray}
where $\vert \phi^{+} \rangle_{A_2 B}$ is one of the Bell states given by Eq. (\ref{bellbasis1}) and the resource state (\ref{EntMix})
is characterised by the mixing parameter $p$. Alice seeks to teleport to Bob the unknown state $\vert \psi_{in}\rangle$ defined in Eq. (\ref{instate}). The joint state of the entire system at $t=0$ is given by
\begin{eqnarray}
\label{jointstate2}
\varrho_{A_1 A_2 B} = \rho_{A_1} \otimes \rho_{A_2 B}^{w},
\end{eqnarray}
where $\rho_{A_1}=\vert \psi_{in}\rangle\langle \psi_{in}\vert$ and $\rho_{A_2 B}^{w}$ are the density operators
corresponding respectively to the state to be teleported and the Werner-type state given by Eq.~(\ref{EntMix}). Then Alice's two
qubits $A_1$ and $A_2$ evolve under a common dephasing environment, while Bob's qubit $B$ evolves under a local
dephasing environment. After noisy time evolutions from $t=0$ to $t=\tau$ in both the wings of Alice and Bob, Alice
performs her Bell-basis measurements at time $t=\tau$. Subsequently, the joint state of the entire system becomes
\begin{eqnarray}
\label{rhowernertype}
\varrho_{A_1 A_2 B}(\tau) &=& \frac{1}{4}\vert \phi^{+} \rangle_{A_1A_2} \langle \phi^{+} \vert \otimes \varrho_B^{\phi^+}(a_{\tau},\:b_{\tau}) + \frac{1}{4}\vert \phi^{-} \rangle_{A_1A_2} \langle \phi^{-} \vert \otimes \varrho_B^{\phi^-}(a_{\tau},\:b_{\tau}) \nonumber \\
&+& \frac{1}{4}\vert \psi^{+} \rangle_{A_1A_2} \langle \psi^{+} \vert \otimes \varrho_B^{\psi^+}(b_{\tau}) + \frac{1}{4}\vert \psi^{-} \rangle_{A_1A_2} \langle \psi^{-} \vert  \otimes \varrho_B^{\phi^-}(b_{\tau}).
\end{eqnarray}
When Alice's BM measurement outcomes correspond to
$\vert \phi^{+} \rangle_{A_1A_2}$ ($\vert \phi^{-} \rangle_{A_1A_2}$), the respective reduced density operators for Bob's qubits are given by
\begin{eqnarray}
\label{wbm1}
\varrho_B^{\phi^+}(a_{\tau},\:b_{\tau}) = p_{\uparrow \uparrow} \vert \uparrow \rangle_{B} \langle \uparrow \vert +
p_{\uparrow \downarrow} \vert \uparrow \rangle_{B} \langle \downarrow \vert
+ p_{\downarrow \uparrow} \vert \downarrow \rangle_{B} \langle \uparrow \vert +
p_{\downarrow \downarrow} \vert \downarrow \rangle_{B} \langle \downarrow \vert, \\
\varrho_B^{\phi^-}(a_{\tau},\:b_{\tau}) = p_{\uparrow \uparrow} \vert \uparrow \rangle_{B} \langle \uparrow \vert -
p_{\uparrow \downarrow} \vert \uparrow \rangle_{B} \langle \downarrow \vert - p_{\downarrow \uparrow}
\vert \downarrow \rangle_{B} \langle \uparrow \vert + p_{\downarrow \downarrow} \vert \downarrow \rangle_{B}
\langle \downarrow \vert.
\label{wbm2}
\end{eqnarray}

\noindent
When Alice obtains the BM outcomes corresponding to $\vert \psi^{+}\rangle_{A_1A_2} (\vert \psi^{-}\rangle_{A_1A_2})$,
the state of the qubit $B$ in Bob's wing becomes respectively
\begin{eqnarray}
\label{wbm3}
\varrho_B^{\psi^+}(b_{\tau}) = q_{\uparrow \uparrow} \vert \uparrow \rangle_{B} \langle \uparrow \vert +
q_{\uparrow \downarrow} \vert \uparrow \rangle_{B} \langle \downarrow \vert
+ q_{\downarrow \uparrow} \vert \downarrow \rangle_{B} \langle \uparrow \vert +
q_{\downarrow \downarrow} \vert \downarrow \rangle_{B} \langle \downarrow \vert, \\
\varrho^{\psi^{-}}_{B}(b_{\tau}) = q_{\uparrow \uparrow} \vert \uparrow \rangle_{B} \langle \uparrow \vert -
q_{\uparrow \downarrow} \vert \uparrow \rangle_{B} \langle \downarrow \vert
- q_{\downarrow \uparrow} \vert \downarrow \rangle_{B} \langle \uparrow \vert +
q_{\downarrow \downarrow} \vert \downarrow \rangle_{B} \langle \downarrow \vert,
\label{wbm4}
\end{eqnarray}
where the above matrix elements of the density operator of Bob's qubit $B$ are given by
\begin{eqnarray}
\nonumber
&& p_{\uparrow \uparrow}\!=\!\frac{1}{2} + \frac{p}{2} \left( \vert \alpha \vert^2 \!-\! \vert \beta \vert^2 \right), ~
p_{\uparrow \downarrow}\!=\!p~\alpha \beta^{\ast}~a_{\tau} b_{\tau} \\
&& p_{\downarrow \uparrow}\!=\!p~\alpha^{\ast} \beta~a_{\tau}^{\ast} b_{\tau}^{\ast}, ~
p_{\downarrow \downarrow}\!=\!\frac{1}{2} - \frac{p}{2} \left(\vert \alpha \vert^2 \!-\! \vert \beta \vert^2 \right)
\label{pmat}
\end{eqnarray}
and
\begin{eqnarray}
 q_{\uparrow \uparrow}=p_{\downarrow \downarrow}, ~~q_{\uparrow \downarrow}= p~ \alpha^{\ast} \beta ~b_{\tau},~~
 q_{\downarrow \uparrow} = p~ \alpha \beta^{\ast} ~b_{\tau}^{\ast}, ~~q_{\downarrow \downarrow}=p_{\uparrow \uparrow}.
\label{qmat}
\end{eqnarray}

\noindent The state given by Eq.(\ref{rhowernertype}) corresponds to an ensemble arising from the Bell measurements (BM).
It is evident from Eqs.~(\ref{wbm1}) and (\ref{wbm2}) that for Alice's BM outcomes corresponding to both
$(\vert \phi^+ \rangle_{A_1 A_2})$ and $(\vert \phi^- \rangle_{A_1 A_2})$, the teleported states in Bob's wing
will embody strong dephasing effects characterised by the decoherence factors of both the wings, $a_{\tau}$ and $b_{\tau}$.
Consequently, knowing about these outcomes from Alice, Bob will have {\it to discard} the corresponding qubits in his
wing $B$ for ensuring higher QT fidelity. \\

\noindent
On the other hand, if Alice's BM outcomes correspond to $\vert \psi^{+} \rangle_{A_1A_2} (\vert \psi^{-} \rangle_{A_1A_2})$,
it is evident from Eqs.~(\ref{wbm3}) and (\ref{wbm4}) that for these two BM outcomes of Alice, Bob’s qubits are unaffected by
dephasing from Alice’s environment. In these two cases, Bob implements the required unitary transformations depending upon the
BM results communicated by Alice. For example, if Alice informs her BM outcome as corresponding to
$\vert \psi^{+} \rangle_{A_1A_2}$, Bob will apply the unitary operation $U_{x}=\sigma_{x}$, transforming the state of
his qubit into the state $\varrho^{\psi^{+}}_{out}(t)=U_{x}~\varrho^{\psi^{+}}_{B}(b_{\tau})~U_{x}^{\dagger}$ given by
\begin{eqnarray}
\varrho_{out}^{\psi^+}(b_{\tau}) = q_{\uparrow \uparrow} \vert \downarrow \rangle_{B} \langle \downarrow \vert  +
q_{\uparrow \downarrow} \vert \downarrow \rangle_{B} \langle \uparrow \vert
+ q_{\downarrow \uparrow} \vert \uparrow \rangle_{B} \langle \downarrow \vert +
q_{\downarrow \downarrow} \vert \uparrow \rangle_{B} \langle \uparrow \vert
\label{wbm5}
\end{eqnarray}
Similarly, for Alice's BM outcome corresponding to $\vert \psi^{-}\rangle_{A_1A_2}$, Bob will apply the unitary
operation $U_{y}=i\sigma_{y}$, transforming his qubit's state to the form
\begin{eqnarray}
\varrho^{\psi^{-}}_{out}(b_{\tau}) = U_{y}~\varrho^{\psi^{-}}_{B}(\tau)~U_{y}^{\dagger} = \varrho^{\psi^{+}}_{out}(b_{\tau}).
\label{Uy3}
\end{eqnarray}

\noindent
Using Eqs.~(\ref{wbm5}) and (\ref{Uy3}), the corresponding \textit{FTS} \cite{oh2002fidelity} is given by
\begin{eqnarray}
\label{mixfid1}
f(p) &=& \langle \psi_{in}\vert \varrho^{\psi^{+}}_{out}(\tau) \vert \psi_{in}\rangle
= \langle \psi_{in} \vert \varrho^{\psi^{-}}_{out}(\tau) \vert \psi_{in}\rangle \\
\nonumber
&=& \vert \alpha \vert^2 ~q_{\downarrow \downarrow} + \alpha \beta^{\ast} ~q_{\uparrow \downarrow}
+ \alpha^{\ast} \beta ~q_{\downarrow \uparrow} + \vert \beta \vert^2 ~q_{\uparrow \uparrow}.
\end{eqnarray}
Substituting the values of $q_{\downarrow \downarrow}$, $q_{\uparrow \downarrow}$,
$q_{\downarrow \uparrow}$, and $q_{\uparrow \uparrow}$ from Eqs.~(\ref{pmat}) and (\ref{qmat}) in
Eq.~(\ref{mixfid1}) we obtain
\begin{eqnarray}
\label{mixfid1a}
f(p) = p~ \vert \alpha \vert^2 \vert \beta \vert^{2} \left( b_{\tau} +  b_{\tau}^{\ast}  \right) + \frac{1}{2}
+ \frac{p}{2} \left( \vert \alpha \vert^{4} +  \vert \beta \vert^{4} - 2 \vert \alpha \vert^2 \vert \beta \vert^{2}  \right).
\label{mixfid2}
\end{eqnarray}
Let us now parametrize the input state $\vert \psi_{in}\rangle$ given by Eq.~(\ref{instate})
as before by taking $\alpha=\cos(\theta/2)$ and $\beta=\sin(\theta/2)e^{i\phi}$.
Then the fidelity of the teleported state given by Eq.~(\ref{mixfid2}) becomes
\begin{eqnarray}
\label{mixfid3}
&& f(p,\theta,\phi) = p \cos^{2}(\theta/2) \sin^{2}(\theta/2)
\left( b_{\tau} +  b_{\tau}^{\ast}  \right) + \frac{1}{2} \\
\nonumber
&& + \frac{p}{2} \Big[ \cos^{4}(\theta/2) + \sin^{4}(\theta/2)
- 2 \cos^{2}(\theta/2) \sin^{2}(\theta/2) \Big].
\end{eqnarray}
The fidelity $f(p,\theta,\phi)$ given by Eq.~(\ref{mixfid3}) depends on the input state parameters $\theta$ and $\phi$
as well as the mixing parameter $p$. Since, in general, the state to be teleported is unknown, we have to estimate the
fidelity by taking an average over the relevant ranges of the parameters specifying the state being teleported. \\

\noindent
Thus, when Alice and Bob share the mixed entangled state $\rho_{A_2B}^{w}$ given by Eq.~(\ref{EntMix}), the
average \textit{FTS} is given by
\begin{eqnarray}
\label{avtelepfidmixed}
{\cal F}_M(b_{\tau})  = \frac{p}{6} ~\Big( b_{\tau} +  b_{\tau}^{\ast} \Big) + \frac{p}{6} + \frac{1}{2}.
\end{eqnarray}

\noindent
Here, the decoherence factor $b_{\tau}$ is given by Eq.~(\ref{deco02}), which is determined by Bob's noise
parameters. Next, we substitute the expression for the decoherence factor $b_{\tau}$ from Eq.~(\ref{deco02})
into Eq.~(\ref{avtelepfidmixed}) to obtain
\begin{eqnarray}
{\cal F}_M(b_{\tau}) = \frac{p}{3} \cos(\omega_0 \tau) ~ e^{-2 \gamma \ln(1+ \Lambda^2 \tau^2)} + \frac{p}{6} + \frac{1}{2}.
\label{mixfid7}
\end{eqnarray}
We can now vary the mixing parameter $p$ of the Werner-type state and study how the above evaluated
average \textit{FTS} changes under Bob's dephasing noise. Note that the concurrence \cite{wootters2001entanglement} of the
mixed Werner-type state (\ref{EntMix}) shared by Alice
and Bob is given by
\begin{eqnarray}
\label{conmixed}
\mathcal C_m =  \frac{3\:p - 1}{2}, \:\:\:\: p > \frac{1}{3}.
\end{eqnarray}
\noindent Using Eq.~(\ref{conmixed}), the average \textit{FTS} given by Eq.~(\ref{mixfid7}) can then be re-expressed as
\begin{eqnarray}
\label{mixfid9}
{\cal F}_M(b_{\tau}) = \frac{2}{9}\Big(\mathcal C_m + \frac{1}{2}\Big) \cos(\omega_0 \tau)
~ e^{-2 \gamma \ln(1+ \Lambda^2 \tau^2)} + \frac{1}{9}\Big(\mathcal C_m + 5\Big).
\end{eqnarray}

\begin{figure}[h]
\centering
\includegraphics[width=\columnwidth]{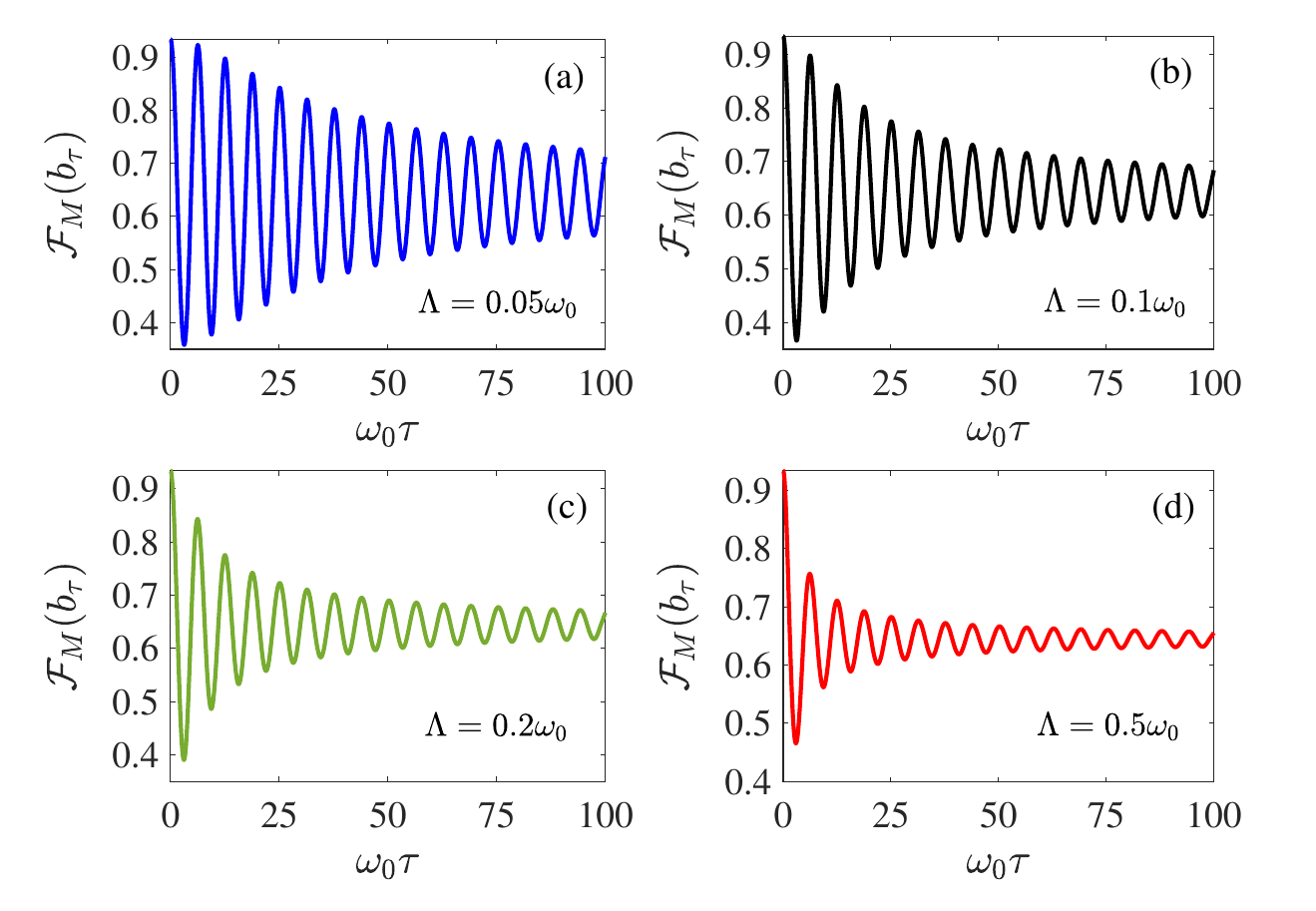}
\caption{\label{mixfid01} We present the average fidelity of the teleported state ${\cal F}_M(b_{\tau})$
as a function of $\tau$-the Bell-basis measurement timing in Alice's wing. The concurrence of the shared
mixed entangled state is fixed at $\mathcal{C}_{m}=0.8$, and the system-environment coupling strength at Bob's
wing is set to $\gamma=0.2\omega_0$. The panels (a)-(d) illustrate the results for different cutoff frequencies pertaining to Bob's environmental model:
(a) $\Lambda=0.05\omega_0$, (b) $\Lambda=0.1\omega_0$, (c) $\Lambda=0.2\omega_0$, and (d) $\Lambda
=0.5\omega_0$, respectively.}
\end{figure}
\begin{figure}[h]
\centering
\includegraphics[width=\columnwidth]{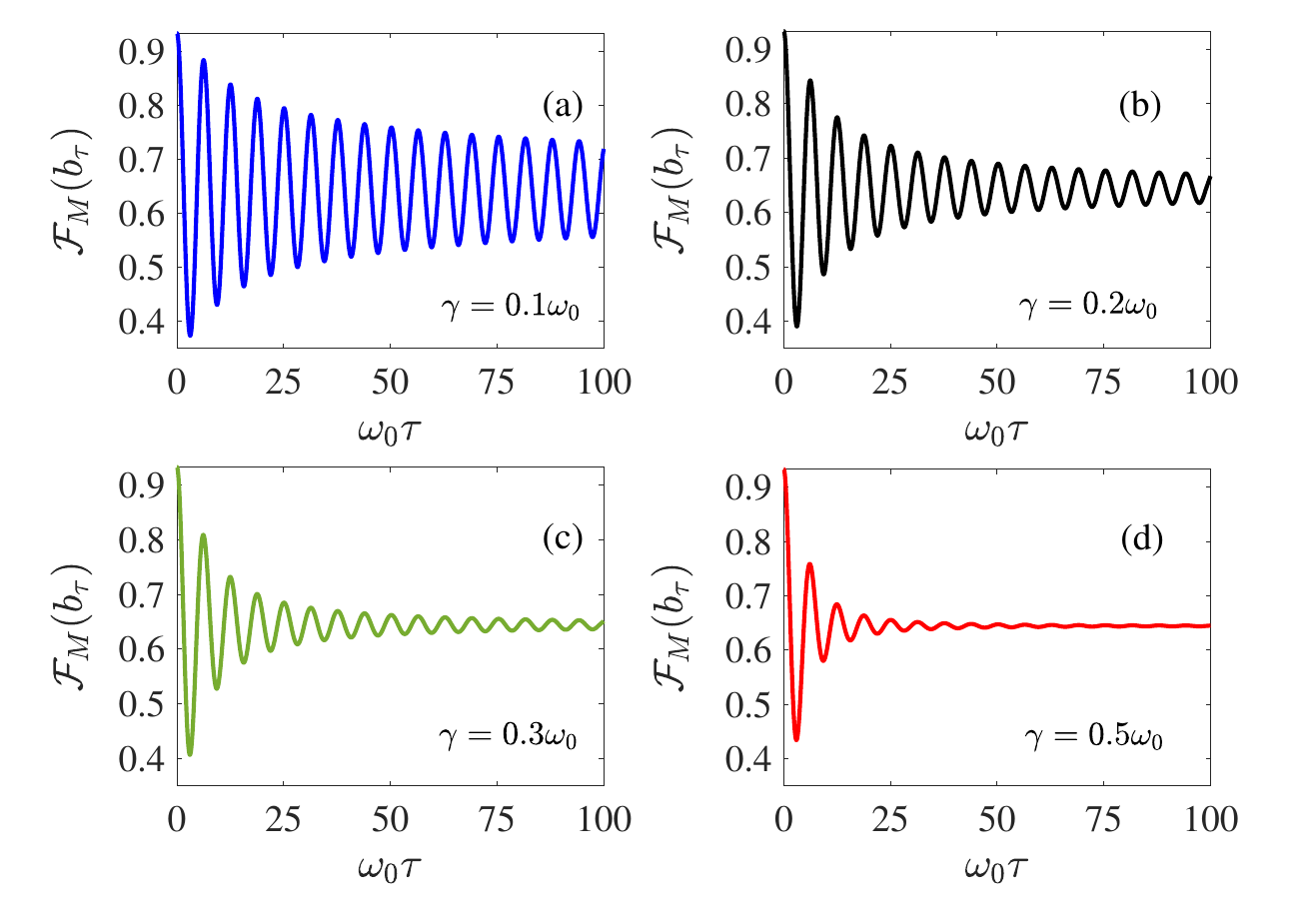}
\caption{\label{mixfid02} We present the average fidelity of the teleported state ${\cal F}_M(b_{\tau})$ as
a function of the Bell-basis measurement timing $\tau$ in Alice’s wing, for different system-environment
coupling strengths $\gamma$. The concurrence of the shared mixed entangled state is fixed at
$\mathcal{C}_{m}=0.8$, and the cutoff frequency of the reservoir is set to $\Lambda=0.2\omega_0$.
The panels (a)-(d) display the results for increasing coupling strengths: (a) $\gamma=0.1\omega_0$, (b)
$\gamma=0.2\omega_0$, (c) $\gamma=0.3\omega_0$, and (d) $\gamma=0.5\omega_0$, respectively.}
\end{figure}

\noindent
In accordance with Eq.~(\ref{mixfid9}), we present in Fig.~\ref{mixfid01} the average fidelity of the
teleported state ${\cal F}_M(b_{\tau})$ as a function of Alice's Bell-basis measurement timing $\tau$ when
Alice and Bob share the mixed entangled state $\rho_{A_2B}^{w}$ given by Eq.~(\ref{EntMix}).
Following our QT protocol, Bob fixes the environmental noise parameters $\Lambda$ and $\gamma$
in his side, and information about the values of these parameters is shared in advance with Alice.
Specifically, the system-environment coupling strength is kept fixed at the value $\gamma=0.2\omega_0$.
In Fig.~\ref{mixfid01}, the cutoff frequencies of Bob's environment are taken as (a) $\Lambda=0.05\omega_0$
(b) $\Lambda=0.1\omega_0$ (c) $\Lambda=0.2\omega_0$ and (d) $\Lambda=0.5\omega_0$ respectively.
Using this pre-shared noise information, Alice optimally selects her measurement time $t=\tau$ to maximize
the average fidelity of the teleported state ${\cal F}_M(b_{\tau})$ given by Eq.~(\ref{mixfid9}).
Note that this average fidelity exhibits damped oscillations with the local maxima points occurring at
$\omega_0\tau=2n\pi$ ($n=1,2,\ldots$). It can be seen from Fig.~\ref{mixfid01} that to achieve
a higher value of average {\it FTS}, Alice is required to perform her Bell measurements within sufficient
short time. In particular, Figure \ref{mixfid01}(a) indicates that the maximum average {\it FTS} can closely
approach $0.9$ when the Bell-basis measurement is carried out within the time $\tau$ satisfying the
condition $0 < \omega_0 \tau < 15$. Further, note that the faster decay of average fidelity of teleported state
is observed in Figs.~\ref{mixfid01}(b), (c), and (d) for higher values of cutoff frequency $\Lambda$ of Bob's
environment. In contrast to Fig.~\ref{mixfid01}, in Fig.~\ref{mixfid02}, we show the average fidelity
${\cal F}_M(b_{\tau})$ as a function of $\tau$ for the same shared mixed entangled state $\rho_{A_2B}^{w}$,
while keeping the cutoff frequency fixed at the value $\Lambda=0.2\omega_0$ and varying the system-environment
coupling strength $\gamma$ across the panels \ref{mixfid02}(a)-\ref{mixfid02}(d).
Guided by the pre-shared information about the noise parameters in Bob's wing, Alice
optimizes the measurement timing $\tau$ towards maximising the value of ${\cal F}_M(b_{\tau})$.
Here an important point is that as $\gamma$ increases, the oscillations in ${\cal F}_M(b_{\tau})$ become
more strongly damped, leading to a faster decay of fidelity and a reduced measurement time window for
achieving higher teleportation fidelity. \\

\begin{table}[h]
\caption{This Table shows the values of the average fidelity for those values of $p$ for which the quantum correlations of the Werner type state do not violate the Bell-CHSH inequality, along with the corresponding maximum values of the Bell-parameter $\langle \mathcal {B}_{max}\rangle^{\rho^{w}}$ and the relevant values of the concurrence $\mathcal {C}_m$ of the state.}\label{tab1}%
\begin{tabular}{@{}llllll@{}}
\toprule
$p$ &  $0.4$ & ~$0.5$ & ~$0.6$ & ~$0.66$ & ~$0.69$\\
\midrule
$\mathcal {C}_m$ &  $0.1$ & ~$0.25$ & ~$0.4$ & ~$0.49$ & ~$0.54$\\
\midrule
$\langle \mathcal {B}_{max}\rangle^{\rho^{w}}$ & $1.13$  & ~$1.41$ & ~$1.69$ & ~$1.87$ & ~$1.95$\\
\midrule
${\cal F}_M(b_{\tau})$ & $0.69$ & ~$0.74$ & ~$0.79$ & ~$0.82$ & ~$0.84$ \\
\midrule
\end{tabular}
\end{table}

\begin{table}[h]
\caption{This Table shows the values of the average fidelity for those values of $p$ for which the quantum correlations of the Werner type state violate the Bell-CHSH inequality, along with the corresponding maximum values of the Bell-parameter $\langle \mathcal {B}_{max}\rangle^{\rho^{w}}$ and the relevant values of the concurrence $\mathcal {C}_m$ of the state.}\label{tab1}%
\begin{tabular}{@{}llllllll@{}}
\toprule
$p$ &  $0.72$ & ~$0.75$ & ~$0.85$ & ~$0.90$ & ~$0.95$\\
\midrule
$\mathcal {C}_m$ & $0.58$ & $0.63$ & $0.78$ & $0.85$ & $0.93$\\
\midrule
$\langle \mathcal {B}_{max}\rangle^{\rho^{w}}$ &  $2.04$ & $2.12$ & $2.40$ & $2.54$ & $2.68$\\
\midrule
${\cal F}_M(b_{\tau})$ & $0.86$ & $0.88$ & $0.93$ & $0.95$ & $0.98$ \\
\midrule
\end{tabular}
\end{table}

\noindent
In Tables $2$ and $3$, we display illustrative results to show how in our scheme the average fidelity ${\cal F}_M(b_{\tau})$
of the teleported state for a Werner-type mixed state used as the resource channel varies with the relevant parameters. The
degree of entanglement of the mixed state, quantified by its concurrence $C_m$, is controlled
through the mixing parameter $p$. On Bob’s side, the noise characteristics are fixed
by choosing the coupling strength between the system and the environment
$\gamma=0.1\omega_0$ and the environmental cutoff frequency $\Lambda=0.02\omega_0$. Alice
performs Bell-basis measurements on her qubits $A_1$ and $A_2$ at an appropriately
chosen time, determined by the given values of Bob’s noise parameters. The measurement time is set to
$\omega_0\tau=2\pi$ pertaining to both the Tables $2$ and $3$. \\

\noindent
Now, there are two cases for the Werner-type state (\ref{EntMix}) as resource with respect to its local/non-local
behavior. The state is local (i.e. the quantum correlations do not violate the Bell-CHSH inequality) as well as
entangled for $\frac{1}{3} < p \leq \frac{1}{\sqrt{2}}$. On the other hand, the state is non-local (i.e. the
quantum correlations violate the Bell-CHSH inequality) and entangled when $ p > \frac{1}{\sqrt{2}}$.
Note that the Werner-type state (\ref{EntMix}) does not violate the Bell-CHSH inequality when
$0 < \mathcal C_m \leq 0.56$, and the state is non-local when $\mathcal C_m > 0.56$. \\

\noindent In Table $2$, we show the average fidelity for certain values of the state parameter $p$ of the Werner type
state for which the Bell-CHSH inequality is {\it not violated} by the quantum correlations, alongside displaying the
corresponding maximum values of the Bell-parameter and the relevant values of concurrence. In Table $3$,
we present the values of average fidelity for those values of the state parameter $p$ for which the Bell-CHSH
inequality is {\it violated} by the quantum correlations, along with the corresponding maximum values of
the Bell-parameter and the relevant values of the concurrence. \\

\section{Methods}

\subsection{The time evolution in Alice's wing}

\noindent The microscopic Hamiltonian of two qubits in a common dephasing environment in  Alice's wing can be described by
\begin{eqnarray}
\label{alicewing}
\!\!\!\!H\!= H_S + H_E + H_{SE}
\end{eqnarray}
where, $H_S =  \frac{\hbar}{2} \sum_{i=1}^2 \!  \omega_0 \sigma_{z}^i$ is the system Hamiltonian representing the two qubits at Alice's side, and $\omega_0$ is the qubit transition frequency between the two energy levels of each qubit, while $\sigma_z^i$ are the Pauli spin operators for the qubits ($i = A_1, A_2)$ in Alice's wing. The environmental Hamiltonian $H_E = \hbar\!\sum_{k}\! \omega_{k} a_{k}^{\dagger} a_{k}$ is the common dephasing environment modeled as a collection of Bosonic field modes with frequencies $\omega_k$ where $a_k^{\dagger}$ and $a_k$ are respectively the creation and annihilation operators of the $k^{th}$ mode of the environment. The two qubits ($A_1,\:A_2$) in Alice's wing interact with the common dephasing environment where the system-environment interaction is described by the Hamiltonian $H_{SE} = \hbar S_{z} \sum_{k} (g_{k} a_{k}^{\dagger} + g_{k}^* a_{k})$. Here the system operator $S_z$ is coupled
to the bath operator $\sum_{k} (g_{k} a_{k}^{\dagger} + g_{k}^* a_{k})$, while the operator $S_z = \sum_i\:\sigma_z^i$ is the
collective spin operator of the two-qubit system in Alice's wing and the parameter $g_k$ denotes the coupling strength between the
system and the $k^{th}$ mode of the environment. The environment is initially taken to be in thermal equilibrium at temperature $T$
and is considered to be initially decoupled from the environment. \\

\noindent
Considering the environment to be initially ($t=0$) in thermal equilibrium at temperature $T$, taken to be decoupled from the
system, one can derive the Liouville-von Neumann master equation for these two-qubits under the above specified common
dephasing noise as
\begin{eqnarray}
\label{liouville-neumann}
\frac{d}{dt} \varrho_{A_1A_2} (t)
= -\frac{i}{\hbar}\Big[H_S, \varrho_{A_1A_2} (t) \Big] + {\mathcal L(t)} \varrho_{A_1A_2} (t),
\end{eqnarray}
where the dissipative term causing the decoherence is given by
\begin{eqnarray}
\nonumber
{\mathcal L(t)} \varrho_{A_1A_2} (t)
&=& \frac{1}{2}\Big( A(t) ~S_z \varrho_{A1A2} (t) S_z - 2\alpha(t) ~S_z S_z \varrho_{A_1A_2} (t) \\
{}&& ~~~~~~~ - 2 \alpha^{\ast}(t) ~\varrho_{A1A2} (t) S_z S_z\Big).
\label{liouville}
\end{eqnarray}
The time dependent functions $A(t)$ and $\alpha(t)$ determining the rate of decoherence
are given by
\begin{eqnarray}
\label{dtalphat}
A(t) &=& 4 \int_{0}^{\infty} J(\omega)\coth\Big(\frac{\hbar \omega}{2k_{B}T}\Big)\frac{\sin(\omega t)}{\omega}d\omega ,
\label{A(t)}
\end{eqnarray}
\begin{eqnarray}
\alpha(t) &=& \frac{A(t)}{4}-i \int_{0}^{\infty} J(\omega)\frac{1-\cos(\omega t)}{\omega}.
\label{alphat}
\end{eqnarray}
Note that $J(\omega) = \sum_k |\: g_k\:|^2\delta(\omega - \omega_k)$ is the spectral density of the dephasing environment.
If the reservoir spectrum is continuous, $g_k \rightarrow g(\omega)$, and consequently we have $J(\omega) = P(\omega)|g(\omega)|^2$,
where $P(\omega)$ is the density of states of the noisy environment. \\

\noindent
Under the quantum master equation (\ref{liouville-neumann}), a two-qubit density matrix in Alice's wing
evolving up to time $t=\tau$ is given by
\begin{eqnarray}
\varrho_{A_1A_2} (\tau) \!\!&=&\!\!
\begin{pmatrix}
\varrho_{11}(0) & f_\tau \varrho_{12}(0) & f_\tau \varrho_{13}(0) & a_\tau \varrho_{14}(0)\\
f^{\ast}_\tau \varrho_{21}(0) & \varrho_{22}(0) & \varrho_{23}(0) & g^{\ast}_\tau \varrho_{24}(0)\\
f^{\ast}_\tau \varrho_{31}(0) & \varrho_{32}(0) & \varrho_{33}(0) & g^{\ast}_\tau \varrho_{34}(0)\\
a^{\ast}_\tau \varrho_{41}(0) & g_\tau \varrho_{42}(0) & g_\tau \varrho_{43}(0) & \varrho_{44}(0)
\end{pmatrix},
\label{densitya1a2}
\end{eqnarray}
where $f_\tau=e^{{-i\omega_0\tau}-4\int_{0}^{\tau} \alpha(t) ~dt}$ and
$g_\tau=e^{{i\omega_0 \tau}-4\int_{0}^{\tau} \alpha(t) ~dt}$ are determined by the function
$\alpha(t)$, whereas the key decoherence factor in our treatment,
$a_\tau=e^{{-2i\omega_0 \tau}-4\int_{0}^{\tau} A(t) ~dt}$ is determined by the function $A(t)$.
The qubits $A_1$ and $A_2$ evolve under the above quantum master equation (\ref{densitya1a2})
up to the time $t=\tau$, when Alice performs her Bell basis measurements.

\subsection{The time evolution in Bob's wing}

\noindent The qubit $B$ in Bob's wing evolves under a local dephasing noise. To describe the local evolution
of the qubit $B$ we consider the spin-boson Hamiltonian describing the local dephasing of a two-level system
in terms of the interaction of a two-level system with a bosonic environment, defined as
\begin{eqnarray}
\label{bobewing}
\!\!\!\!{\cal H}\!= {\cal H}_S + {\cal H}_E + {\cal H}_{SE},
\end{eqnarray}
where ${\cal H}_S=\frac{\hbar}{2}\omega_{0} \sigma_{z}$ is the system Hamiltonian representing the single qubit in Bob's side.
The environmental Hamiltonian is ${\cal H}_E = \hbar \sum_{k}  \omega_{k}b_{k}^{\dagger}b_{k}$. The local dephasing environment
in Bob's wing is modeled by a collection of harmonic oscillators with frequencies $\omega_k$ where $b_{k}^{\dagger}$ and $b_{k}$
are the creation and annihilation operators pertaining to the $k^{th}$ mode of Bob's environment. The qubit in Bob's wing interacts with its local
environment, which is modelled by the Hamiltonian ${\cal H}_{SE} = \hbar \sigma_{z} \sum_{k}
( v_{k}  b_{k}^{\dagger} + v_{k}^{\ast} b_{k} )$. As before, $\sigma_z$ is the observable associated to the system and $\sum_{k}
( v_{k}  b_{k}^{\dagger} + v_{k}^{\ast} b_{k})$ is the operator corresponding to the environment, where the coupling strength
between Bob's qubit and the $k^{th}$ mode of the environment is denoted by the parameter $v_{k}$. The parameter $v_{k}$ can,
in general, be different from the coupling strength $g_k$ in Alice's wing. The environment is initially ($t=0$) considered in
thermal equilibrium at temperature $T$, decoupled from Bob's qubit. Subjected to this local dephasing environment, Bob's qubit
evolves according to the Liouville-von Neumann master equation
\begin{eqnarray}
\label{masterB}
\!\!\!\!\!\!\!\frac{d}{dt} \rho_{B}(t)
\!=\! -\frac{i}{\hbar}\Big[{\cal H}_S, \rho_{B} (t) \Big] \!+\! \frac{B(t)}{2} \Big(\sigma_{z} \rho_{B} (t) \sigma_{z}
\!-\! \rho_{B} (t) \Big).
\end{eqnarray}
The time-dependent function $B(t)$ determining the decoherence rate is
\begin{eqnarray}
\label{decoB}
B(t)=4 \int_{0}^{\infty} {\cal J} (\omega)\coth\Big(\frac{\hbar \omega}{2k_{B}T}\Big)\frac{\sin(\omega t)}{\omega}d\omega.
\end{eqnarray}
Here ${\cal J} (\omega) = \sum_k |\: v_k\:|^2\delta(\omega - \omega_k)$ is the spectral density of Bob's environment,
which can, in general, be different from that of Alice's environment. If the environmental spectral density is continuous,
$v_k \rightarrow v(\omega)$, then we have ${\cal J} (\omega)={\cal P}(\omega)|v(\omega)|^2$,
where ${\cal P}(\omega)$ is the density of states of Bob's environment. According to this quantum master equation
(\ref{masterB}), a single-qubit density matrix in Bob's wing evolves as
\cite{goan2010non,chen2014investigating,addis2014coherence}
\begin{eqnarray}
\label{eq10}
\rho_{B}(\tau) &=& \begin{pmatrix}
\rho_{11} (0) &  b_\tau \rho_{12} (0) \\
b_\tau^{\ast}  \rho_{21} (0)  & \rho_{22} (0) \\
\end{pmatrix},
\end{eqnarray} \\
where the decoherence factor $b_\tau=e^{{-i\omega_0 \tau}-\int_{0}^{\tau} B(t) ~dt}$ is determined by the function $B(t)$
which depends on the noise parameters of Bob's environment. \\

\subsection{Wootters' concurrence formula}

\noindent We recall Wootter's formula for concurrence \cite{wootters1998entanglement}, which states that for a bipartite state $\rho$, the concurrence ($\mathcal{C}$) is defined as
\begin{eqnarray}
\label{concwoot}
\mathcal{C} = \max \lbrace 0, \sqrt{\lambda_1} - \sqrt{\lambda_2} - \sqrt{\lambda_3} - \sqrt{\lambda_4}\rbrace,
\end{eqnarray}
where $\lambda_i$'s are the eigenvalues of $\rho\:\tilde{\rho}$ with $\tilde{\rho}=(\sigma_{y}\otimes \sigma_{y})\rho(\sigma_{y}\otimes \sigma_{y})$ being the spin-flippled density matrix of $\rho$. Now, for the
non-maximally entangled pure resource state shared by Alice and Bob (as defined in Eq.(\ref{EntPure}),
i.e. $\vert \chi\rangle_{A_2 B}$), the Wootters' concurrence is calculated as
\begin{eqnarray}
\label{conpure}
\mathcal C_p = \mathcal{C}(\vert \chi\rangle_{A_2 B})= 2\mu \lambda.
\end{eqnarray}

\subsection{Bell inequality violation}

Apart from relating FTS with the entanglement measure, we have also looked into relating FTS with the Bell-CHSH
violation  of the shared resource state between Alice and Bob signifying nonlocality of the resource state.
\noindent An entangled two-qubit state $\rho$ violates the Bell-CHSH inequality if and only if the following condition is satisfied \cite{horodecki1995violating}:
\begin{eqnarray}
\label{bellchsh}
\mathcal {M}(\rho) = \max_{i>j}(u_{i}+u_{j})>1
\end{eqnarray}
where $u_{i}$'s are the eigenvalues of the matrix $T^{\dagger}T$, with the elements of the matrix $T$ being defined as
\begin{eqnarray}
\label{teleportationfidelity1}
T_{ij} = Tr(\rho ~\sigma_{i}\otimes \sigma_{j}),
\end{eqnarray}
where  $\sigma_{i}$'s are the Pauli matrices. The maximum value of the Bell-CHSH expression for the state $\rho$ is denoted by
$\langle \mathcal {B}_{max}\rangle^{\rho}$ and is given by
\begin{eqnarray}
\label{bmax}
\langle \mathcal {B}_{max}\rangle^{\rho} = 2\sqrt{\mathcal {M}(\rho)}.
\end{eqnarray}
Using Eq.~(\ref{bellchsh}), the values of the quantity $\mathcal{M}({\rho})$ for the shared pure and mixed entangled states
given by Eqs.~(\ref{EntPure}) and (\ref{EntMix}) respectively are
\begin{eqnarray}
\mathcal {M}(\vert \chi \rangle_{A_2 B}) = (\mu + \lambda)^2, ~{\rm and}~~
\mathcal {M}(\rho_{A_2 B}^{w}) =  2~p^2.
\end{eqnarray}

\section{Conclusion}

\noindent In the present work, instead of attempting an incremental improvement in the usual schemes for counteracting the
effects of noise, we have sought to devise an entirely different strategy based on suitably timing Alice's Bell-basis measurement
in sync with the designing of Bob's environmental noise structure. This is implemented by using an appropriate postselection in
Bob's wing so that the fidelity of the state teleported to Bob can be maximised in a way independent of local noise in Alice's wing.
The postselection of Bob's qubit states invoked for this purpose correspond to the two particular outcomes of Bell-basis
measurement that do not entail any noisy effect induced by Alice's environment on the reduced states of Bob's qubit.
On the other hand, for the other two Bell-basis measurement outcomes that entail cumulative effect of noises from both
the environments of Alice and Bob, Bob's qubits are discarded. Another key element of our proposed scheme is the use
of a general non-Markovian spin-boson model adapted for both the environments in the two wings. Hence, from the foundational
point of view, a deeper analysis of the way reservoir memory plays a critical role in achieving the high fidelity of noisy
QT in our scheme could be instructive in providing clues for further variants of our protocol. This is currently being investigated. \\

\noindent Here it is worth mentioning that while, in principle, the formulation of this scheme holds for any two-level quantum system,
its near-term experimental realisability merits closer investigation by using, in particular, the existing optical platforms. For example,
for the photonic systems, in Bob's wing, the required controllable dephasing noise for an Ohmic spectral profile can possibly be
engineered by coupling the polarization and frequency degrees of freedom using birefringent quartz plates \cite{laine2012nonlocal}.
Taking clues from the earlier related experiments \cite{liu2011experimental}, it also seems that the dephasing environment common
to the qubits in Alice's wing can as well be simulated effectively by photons propagating through the birefringent media. Thus, using
the well-established photonic architectures, it should be worth trying to set up the required experimental framework for implementing
our proposed protocol based on Alice's judicious timing of Bell-basis measurement. As explained earlier, such timing needs to be
commensurate with Bob's choice of the relevant parameters characterising his local environment. Then, followed by an appropriate
postselection of Bob's qubit states, maximisation of the teleportation fidelity can be achieved, whatever be the nature and amount
of noise in Alice's wing. In conclusion, it is worth remarking that such a uniquely conceived noise-mitigating protocol, along with its
empirical vindication, is expected to stimulate efforts towards its further adaptations beyond QT in the broader context of open-system
quantum communication protocols such as entanglement swapping and quantum repeaters. Thus, these studies have the potential for
significant applications towards engineering noise-resilient quantum networking protocols operating in suitably structured non-Markovian
environments.

\vskip 0.5cm
\section*{Acknowledgement:} MMA and DH are grateful for helpful discussions with Urbasi Sinha and her group while this work
was being developed. DH acknowledges support from NASI Senior Scientist Fellowship towards initiating this work at Bose Institute, Kolkata.
For completing this work, DH acknowledges the support as Distinguished Visitor to the Light and Matter Physics Group, RRI,
Bengaluru under the National Quantum Mission of the DST, Govt. of India. The authors thank Tridhara Roy for her help in drawing Fig.~1.

\bibliography{RefsNoisyTeleport}

\end{document}